\documentclass[a4paper,11pt]{article}
\usepackage{jheppub} 

\usepackage[T1]{fontenc} 
\usepackage{MnSymbol}
\usepackage{dsfont}
\usepackage{tikz}

\def\tr{\,\mathrm{tr}\,}	
\newcommand{\pard}[2]{\frac{\partial#1}{\partial#2}}
\newcommand{\deriv}[2]{\frac{\mathrm{d}#1}{\mathrm{d}#2}}
\newcommand{\contour}[2]{\oint_{#2} \frac{d#1}{2\pi i}\;}
\def\W{\mathcal{W}}
\def\WA{W_{\mathcal{A}}}

\newcommand{\vprime}{\frac{V'(\omega)}{p-\omega}}

\def\z{\zeta}

\def\d{\delta}
\newcommand{\ev}[1]{\left\langle #1 \right\rangle}

\title{\boldmath Antisymmetric Wilson loops in $\mathcal{N}=4$ SYM beyond the planar limit}

\author[a,b]{James Gordon}

\affiliation[a]{Department of Physics and Astronomy, University of British Columbia,  Vancouver, BC Canada V6T 1Z1}
\affiliation[b]{Nordita, KTH Royal Institute of Technology and Stockholm University,
Roslagstullsbacken 23, SE-106 91 Stockholm, Sweden}

\emailAdd{jbgordon@phas.ubc.ca}

\abstract{We study the $\frac{1}{2}$-BPS circular Wilson loop in the totally antisymmetric representation of the gauge group in $\mathcal N =4 $ supersymmetric Yang-Mills. This observable is captured by a Gaussian matrix model with appropriate insertion. We compute the first $1/N$ correction at leading order in 't Hooft coupling by means of the matrix model loop equations. Disagreement with the 1-loop effective action of the holographically dual $D5$-brane suggests the need to account for gravitational backreaction on the string theory side.}

\begin{document} 
\maketitle
\flushbottom

\section{Introduction}
\label{sec:intro}

Since its inception, the AdS/CFT correspondence has held out the promise of a fully non-perturbative definition of quantum string theory in non-trivial backgrounds. Testing this strongest form of the conjecture is, however, very hard. Progress can be made in this direction by considering controlled deviations from the large-$N$, large-$\lambda$ limit. In an exciting development, the techniques of supersymmetric localization and integrability have in recent years generated a profusion of exact gauge theoretic results, enabling such quantitative testing and ushering in an era of precision holography.

Supersymmetric localization reduces the partition function of $\mathcal N =4$ super-Yang-Mills to a gaussian Hermitian matrix model \cite{PestunLocalizationOfGaugeTheory}. Furthermore, a certain supersymmetry-preserving sub-sector of the theory is completely captured, for arbitrary $N$ and $\lambda$, by matrix model expectation values of appropriate insertions (see eqs. (\ref{eq:HermitianMatrixModel},\ref{eq:MatrixModelInsertion}) in the next section). For the fundamental Wilson loop this was anticipated in the prescient work \cite{Erickson2000} (also \cite{Drukker2001}) where an infinite class of planar diagrams was explicitly re-summed, generating what was correctly conjectured to be the exact planar result. Localization formul\ae\, for more general correlators and Wilson loops have since followed.

In the study of non-abelian gauge theories, the Wilson loop operator plays a fundamental role. In addition to serving as an order parameter for the confinement-deconfinement phase transition, it provides a naturally gauge-invariant formulation of the theory which, while inherently non-local, is quite natural from the point of view of the correspondence to string theory. It can be understood as the phase acquired by a probe particle as it traces out some closed path $C$. As well as this contour, the Wilson loop is labeled by a representation $\mathcal R$ of the gauge group, describing the charge of the probe particle. In $\mathcal N = 4$ SYM, the natural (supersymmetric and UV finite) Wilson loop observable also includes a scalar field coupling:
\begin{equation} \label{eq:WilsonLoop}
	W(C) \equiv \frac{1}{N} \textrm{Tr}_\mathcal{R} \left( \mathcal{P} \, \exp \left\{ \oint_C \! d\tau ( i A_\mu \dot{x}^\mu + |\dot x | n^I \Phi_I )  \right\} \right)
\end{equation}
For the special case where $C$ is a circle and $n^I$ a constant unit vector, this preserves half of the supersymmetries, permitting its localization after compactification on $S^4$.

To date the correspondence has withstood over two decades of sustained scrutiny. It is therefore noteworthy when tension, let alone disagreement, is found between putatively dual quantities. We can expect such cases to reveal important subtleties or misunderstandings of the dictionary, or indeed to elucidate the limits of its applicability. 

In this paper we study an as-yet unresolved mismatch in the most scrutinized example of AdS/CFT, namely the duality between $\mathcal{N}=4$ supersymmetric Yang-Mills theory with gauge group $SU(N)$, and type IIB superstring theory on $AdS_5\times S^5$. The discrepancy occurs in the 1-loop correction to the $\frac{1}{2}$-BPS circular Wilson loop in the rank-$k$ totally-antisymmetric representation of the gauge group, with $k\sim \mathcal{O} (N)$. We compute this quantity on the gauge theory side by solving the loop equations for the corresponding matrix model obtained from localization\footnote{
A similar approach to deriving Wilson loops in higher representations from the loop equations has been advocated for example in \cite{Akemann2001} where general $k$-loop (fundamental) correlators, from which general representations can be constructed, were computed building on the work of \cite{Ambjoern1993}. However, their results are not directly applicable here as we will be interested in the limit $k\rightarrow \infty$, with $k/N$ fixed.
}. In fact, the antisymmetric Wilson loop was evaluated exactly using orthogonal polynomials in \cite{Fiol2014a}; however, it is not clear how to extract from their result the $1/N$ expansion, which is needed for comparison with holography. Their formula will however be useful for numerical verification of our result.

Apart from its intrinsic interest, an understanding of the structure of higher-rank Wilson loops may also yield insight into the analogous, longstanding $1$-loop matching problem for the fundamental Wilson loop \cite{Forste1999a,Drukker2000a,Sakaguchi2008,Kruczenski2008,Kristjansen2012}. See \cite{Forini2016,Faraggi2016,Forini2017} for recent progress on this problem.

According to the AdS/CFT dictionary, the antisymmetric Wilson loop is dual to a probe $D5$-brane with $k$ units of electric flux on its $AdS_2 \times S^4$ worldvolume \cite{Gomis2006,Gomis2007}, which ``pinches off'' along the circular contour described by the Wilson loop at the boundary of $AdS$. At leading order in large $N$ and $\lambda$ the on-shell action of the $D$-brane was successfully matched with the gauge theory \cite{Yamaguchi2006,HartnollKumar2006}. 

The $D$-brane tension is of order $N$, and its 1-loop effective action captures non-planar contributions. The spectrum of fluctuations and 1-loop effective action were derived in \cite{Faraggi2011,Harrison2012} and \cite{Faraggi2012OneLoopEffectiveActionAntisymmetricWL} respectively, with the result\footnote{The answer of $\frac{1}{12} \ln \sin \theta_k$ quoted in \cite{Faraggi2012OneLoopEffectiveActionAntisymmetricWL} was updated by the authors in the subsequent publication \cite{PandozayasOneLoopStructureWL} to incorporate a missing normalization factor.}
\begin{equation} \label{eq:OneLoopEffectiveAction}
	\Gamma_1 = \frac{1}{6} \ln \sin \theta_k,
\end{equation}
where $\theta_k$ is defined by $(\theta_k -\sin\theta_k \cos\theta_k) = \pi k/N$. A first step towards reproducing this from the gauge theory side was taken in \cite{PandozayasOneLoopStructureWL}. They obtained the same functional dependence on $k$, but a different overall constant:
\begin{equation}
	\tilde \Gamma_1 = \frac{1}{2} \ln \sin \theta_k 
\end{equation}
The mismatch is not surprising, as the computation  neglected the backreaction of the Wilson loop insertion on the equilibrium eigenvalue distribution of the matrix model. Here we will systematically take this into account. However, our result, which withstands convincing numerical testing, still does not match with \eqref{eq:OneLoopEffectiveAction}; even the power of $\lambda$ is different. As we mention in the conclusions, this contribution to the free energy likely corresponds to the gravitational backreaction of the probe $D$-brane, i.e. our result is of very different origin to \eqref{eq:OneLoopEffectiveAction}. It would still be interesting to try to match \eqref{eq:OneLoopEffectiveAction} with a gravity calculation by a careful determination of strong-coupling corrections on both sides.

The layout of the paper is as follows. In section \ref{sec:wilsonloop} we summarize the localization result \cite{PestunLocalizationOfGaugeTheory} for the Wilson loop, and set up the problem. In section \ref{sec:mainsection} we derive a sequence of loop equations for the gaussian matrix model perturbed by the Wilson loop insertion, and solve them for the resolvent up to the second sub-leading order. We then derive from this the free energy, by two different means. We also calculate the correction to this result due to considering gauge group $SU(N)$ instead of $U(N)$.  Section \ref{sec:numerics} presents some numerical checks of our answer, by comparing to the exact result of \cite{Fiol2014a}. Finally we end with some conclusions and open questions in section \ref{sec:Conclusions}.



\section{Antisymmetric circular Wilson loop} \label{sec:wilsonloop}

Localization of $\mathcal N = 4$ SYM reduces the full partition function to that of a Hermitian Gaussian matrix model \cite{PestunLocalizationOfGaugeTheory}
\begin{equation} \label{eq:HermitianMatrixModel}
	Z_{Gauss} = \int [dM] \, e^{-\frac{2N}{\lambda} \tr M^2 } ,
\end{equation}
while the expectation value of the circular Wilson loop is mapped to an expectation value (denoted $\ev{}_0$) in this matrix model:
\begin{equation} \label{eq:MatrixModelInsertion}
	\ev{W_\mathcal{R} (\text{Circle})} = \frac{1}{\text{dim}[\mathcal R]}\ev{\mathrm{tr}_\mathcal{R} \, e^M}_0
\end{equation}
The representation $\mathcal R$ of the gauge group is completely arbitrary at this stage. We will be interested in $\mathcal R = \mathcal{A}_k$, the totally anti-symmetric representation of rank $k$, in the large-$N$, large-$\lambda$ regime with
\begin{equation}
	f \equiv \frac{k}{N} \sim \mathcal O (1)
\end{equation}
held fixed.
The generating function for the character of this representation is
\begin{equation} \label{eq:generatingfn}
	F_{\mathcal{A}}(t) = \det (t+e^M) = \sum_{k=0}^N t^{N-k} \binom{N}{k} \, \WA
\end{equation}
so that we can write the Wilson loop expectation value as \cite{HartnollKumar2006}
\begin{eqnarray} \label{eq:Wcontour}
	\ev{\WA} &=& d_A^{-1} \contour{t}{D} \frac{\ev{F_A(t)}_0}{t^{N-k+1}} ,
\end{eqnarray}
where $D$ encircles the origin and $d_A = \binom{N}{k}$ is the dimension of the representation. The following change of variables, which maps the complex $t$-plane to the cylinder, will prove convenient:
\begin{equation}
	t = e^z .
\end{equation}
It will also be useful to view the expectation value of $F_A$ as defining a family of perturbed partition functions parametrized by $z$,
\begin{equation} \label{eq:ModifiedMatrixModel}
	\mathcal Z (z) \equiv \int [dM] \exp\left\{-\frac{2N}{\lambda} \tr M^2 + \tr \log (1+e^{M-z}) \right\} .
\end{equation}
and to define a corresponding ``free energy''
\begin{equation} \label{eq:Fz}
	\mathcal F(z) \equiv -\frac{1}{N} \log \left[ \mathcal Z (z) Z_{Gauss}^{-1} \right] .
\end{equation}
Note the unconventional factor of $Z_{Gauss}^{-1}$ here. In this manner we obtain the following exact expression for the Wilson loop:
\begin{equation} \label{eq:WLcontour}
	\ev{\WA} = d_A^{-1} \contour{z}{} \exp\left\{N(fz-\mathcal{F}(z))\right\}
\end{equation}

The free energy of the purely gaussian matrix model is $\mathcal O (N^2)$ and has a genus expansion in powers of $1/N^2$, ie. 
\begin{equation}
	-\log Z_{Gauss} = \sum_{n=0,2,4,\ldots} N^{2-n} \mathcal{F}_{Gauss,n} \, ; \qquad \mathcal F_{Gauss,n} \sim \mathcal O (1) 
\end{equation}
Since $\mathcal Z(z)$ differs from $Z_0$ by a perturbation to the action of $\mathcal O(N)$, its logarithm goes in powers of $1/N$, with leading term identical to that of $\log Z_{Gauss}$.
Consequently, $\mathcal{F}(z)$ defined in \eqref{eq:Fz} is $\mathcal{O}(1)$, and we write
\begin{equation} \label{eq:Fexpansion}
	\mathcal{F}(z) = \mathcal{F}_0(z) + \frac{1}{N} \mathcal{F}_1(z) + \frac{1}{N^2} \mathcal{F}_2(z) + \ldots;
    \qquad \mathcal{F}_i(z) \sim \mathcal O(N^0)
\end{equation}
As the exponent in \eqref{eq:WLcontour} is $\mathcal{O} (N)$ we can evaluate the $z$-integral in the saddle-point approximation, which yields
\begin{equation}
	\ev{\WA}  = \frac{d_A^{-1}}{2\pi i} \, e^{N(fz_* - \mathcal F(z_*) ) } \cdot i \left[\frac{2\pi}{N\left|\mathcal{F}''(z_*)\right|}\right]^{\frac{1}{2}} \left( 1 + \mathcal O \left( \frac{1}{N} \right) \right)
\end{equation}
where $z_*$ solves the saddle-point equation $\mathcal F_0'(z_*) = f$.
To the order in $N$ given here, there is no backreaction on the saddle due to $\mathcal{F}_1(z)$. The $i$ prefactor is from analytic continuation of the ``wrong-sign'' quadratic form. Finally, plugging in \eqref{eq:Fexpansion}, we have
\begin{equation} \label{eq:logWexpansion}
	\log \ev{\WA} = N \Big[ fz_* - \mathcal F_0(z_*) \Big]
    +\Big[ - \mathcal{F}_1(z_*) - \frac{1}{2} \log \mathcal{F}_0''(z_*) - \log \left( d_A \sqrt{2\pi N} \right) \Big] + \mathcal{O}(1/N)
\end{equation}
The leading order result was already obtained in \cite{Yamaguchi2006,HartnollKumar2006} and agrees perfectly with the $D5$-brane on-shell action \cite{Yamaguchi2006}. The $\log \mathcal F_0''$ term was obtained in \cite{PandozayasOneLoopStructureWL}. Interestingly, the latter turns out to have the same functional dependence on $k$ as the 1-loop effective action of the $D$-brane, as computed in \cite{Faraggi2012OneLoopEffectiveActionAntisymmetricWL}, but with a different numerical coefficient. One might anticipate that $\mathcal F_1(z_*)$, at leading order in $1/\lambda$, should give a similar contribution, so as to correct the numerical mismatch. In fact this turns out not to be the case: the log term is subleading in $\lambda$.

We review the planar solution in the next subsection. What then remains is to compute the non-planar free energy $\mathcal F_1(z)$ of the matrix model \eqref{eq:ModifiedMatrixModel}. We do this in section \ref{sec:mainsection} by calculating the resolvent $\W(p) \equiv \ev{\tr\frac{1}{p-M}}$ order-by-order in $N$ using the loop equation method. With the resolvent in hand, the free energy can be determined in either of two ways. 
\begin{enumerate}
\item $\lambda$-integral: $\W(p)$ is the generating function of monomial expectation values, $\ev{M^j} = N \oint \!dp \,p^j \W(p)$. But $\ev{M^2}$ is also just the derivative of $\log \mathcal Z$ with respect to $\lambda^{-1}$. Therefore $\mathcal{F}$ is obtained from the resolvent by an integral over $p$ and $\lambda$.
\item Eigenvalue density: At large $N$ the eigenvalues condense into a continuous distribution $\rho(x)$. The $\mathcal{O} (1)$ perturbation to $\rho$ due to the Wilson loop insertion is encoded in the discontinuity of $\W_1(p)$ across its single cut. Fluctuations around the large-$N$ saddle-point of the $\int[dM]$ integral are not needed as they cancel against the same contribution coming from $Z_{Gauss}$.
\end{enumerate}
Naturally, we find exact agreement between these methods.

\subsection{Planar approximation}
\begin{figure}[tbp]
\centering
\includegraphics[width=.49\textwidth]{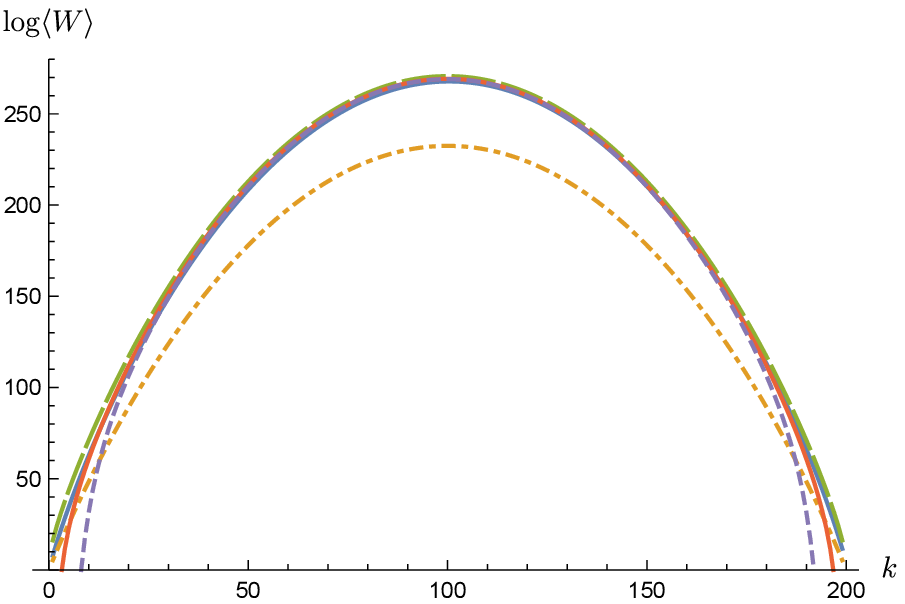}
\hfill
\includegraphics[width=0.49\textwidth]{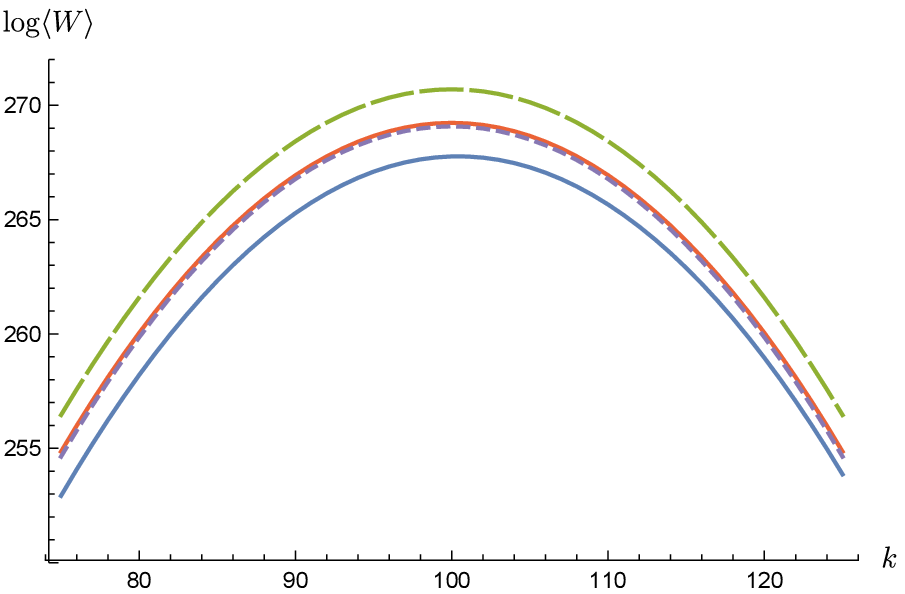}
\caption{\label{fig:strongcoupling} Strong coupling expansion of $\left(\log W \right)_{planar}$ versus $\left(\log W\right)_{exact}$, for $N=200$, $\lambda=35$. The right hand plot is a close-up of the middle region. The solid blue is the exact result $\log W$ evaluated numerically (see section \ref{sec:numerics}). The orange, green, red and purple lines show, in order, the successive approximations to $\left(\log W \right)_{planar}$ given by \eqref{eq:planarexpansion}. They clearly converge to a fixed residual with respect to the exact result, and this residual should be well approximated by the second square bracket in \eqref{eq:logWexpansion}. We confirm this in section \ref{sec:numerics}. In this and subsequent plots we omit the constant prefactor $d_A$.}
\end{figure}

Written in terms of eigenvalues the matrix integral \eqref{eq:ModifiedMatrixModel} is
\begin{equation}
	\mathcal{Z}(z) = \int (\prod_i dm_i) \, e^{-N^2 S[m_i; z]}
\end{equation}
where $S = S_0 + \frac{1}{N} S_1$ and
\begin{subequations} \label{eq:action}
\begin{eqnarray}
	S_0 &=& \frac{2}{\lambda N} \sum_i m_i^2 - \frac{2}{N^2} \sum_{j=1}^N \sum_{i=1}^{j-1} \log |m_i-m_j| \\ \label{eq:action:1}
    S_1 &=& - \frac{1}{N} \sum_i \log (1+e^{m_i -z} ) \label{eq:action:2}
\end{eqnarray}
\end{subequations}
The double sum in $S_0$ is the usual Vandermonde determinant. Recall that at large $N$, the eigenvalues $m_i$ condense into a continuum distribution described by a spectral density $\rho$,
\begin{equation}
	\rho(x) \equiv \frac{1}{N} \sum_i^{N} \delta (x- m_i) ,
\end{equation}
with support $(-\sqrt \lambda,\sqrt \lambda) \subset \mathbb{R}$. This is normalized to unity, and has an expansion
\begin{equation} \label{expansionofrho}
	\rho(x) = \sum_{n=0}^\infty N^{-n} \rho_n(x); \qquad \int_a^b\rho_n(x) dx = \delta_{n0}
\end{equation}
where $\rho_0(x)$ is just the Wigner semicircle distribution
\begin{equation}
	\rho_0(x) = \frac{2}{\pi \lambda} \sqrt{\lambda-x^2}, \qquad -\sqrt{\lambda} < x < \sqrt \lambda,
\end{equation}
since at $N=\infty$ the Wilson loop insertion does not backreact on the eigenvalues. Thus the expectation value of the generating function reduces to its average with respect to $\rho_0(x)$, hence
\begin{equation} \label{eq:planarresult}
	\ev{\WA} \simeq \frac{\sqrt \lambda}{d_A} \contour{\tilde z}{} \exp\left\{ N\left( f\sqrt \lambda \tilde z + \int_{-1}^{1}  dx \rho_0(x) \log (1+e^{\sqrt \lambda(x-\tilde z)})\right)\right\} .
\end{equation}
To facilitate the strong coupling expansion we have re-scaled $z$ according to $z \equiv \sqrt \lambda \tilde z $. From now on we will drop the tilde.
This integral was evaluated in \cite{HartnollKumar2006}, to leading order in large-$\lambda$, using a saddle-point approximation. There is a single saddle-point $z_*\in(-1,1)$ on the real axis, determined by the equation
\begin{subequations}  \label{eq:saddlepointeqn}
\begin{equation}  \label{eq:saddlepointeqn:1}
	\arccos(z_*) - z_* \sqrt{1-z_*^2} = \pi f
\end{equation}
or in an angular parametrization defined by $z_* = \cos \theta_k$:
\begin{equation}  \label{eq:saddlepointeqn:2}
	(\theta_k -\sin\theta_k \cos\theta_k) = \pi f
\end{equation}
\end{subequations}
The Wilson loop \eqref{eq:planarresult} then evaluates to 
\begin{equation}
	\left( \log W_\mathcal{A} \right)_{\textrm{planar}} =  \frac{2N}{3\pi} \sqrt\lambda \sin^3 \theta_k
\end{equation}
This coincides with the on-shell action of the dual $D5$-brane \cite{Yamaguchi2006}. Subsequent terms in the strong coupling expansion are obtained by expanding the logarithm in \eqref{eq:planarresult}, which is like the anti-derivative of the Fermi-Dirac distribution, in inverse powers of $\lambda$ \cite{Horikoshi2016} - see section \ref{sec:numerics}. In figure \ref{fig:strongcoupling} we compare the strong-coupling expansion of the planar approximation obtained in this way with the exact numerical result.

\section{\texorpdfstring{$\boldsymbol{1/N}$}{1/N} expansion from the loop equations} \label{sec:mainsection}
We now set up a systematic expansion around the $N=\infty$, 1-cut solution of the matrix model, using the well-known loop equation approach \cite{Ambjoern1993}. The  matrix model \eqref{eq:ModifiedMatrixModel} is rather exotic - it involves a non-polynomial $\mathcal O(1/N)$ perturbation to the Gaussian potential. Consequently the usual genus expansion familiar from the study of polynomial-potential matrix models (see eg. \cite{Ambjoern1993}) becomes an expansion in $1/N$.

We begin with a few definitions. Our main object of study will be the resolvent, defined by
\begin{subequations}
\begin{eqnarray} \label{eq:resolvent}
	\W (p) &\equiv & \frac{1}{N} \left\llangle \tr \frac{1}{p-M} \right\rrangle \label{eq:resolvent:1}\\ 
    &=& \sum_{n=0}^\infty \frac{1}{N^n} \W_n(p) . \label{eq:resolvent:2}
\end{eqnarray}
\end{subequations}
The double angle-brackets mean the expectation value is with respect to $\mathcal Z(z)$, ie. such expectation values are always functions of $z$. More generally, the ``$s$-loop correlator'' is defined as
\begin{equation} 
	\W(p_1,\ldots,p_s) = N^{s-2} \left\llangle \tr \frac{1}{p_1-M} \ldots \tr \frac{1}{p_s-M} \right\rrangle_{connected}
\end{equation}

The so-called loop equation for the resolvent follows from invariance of the partition function under the infinitesimal change of variables
\begin{equation}
	M\rightarrow M+\epsilon \frac{1}{p-M} \,. 
\end{equation}
The Jacobian for this transformation is $\left( \tr \frac{1}{p-M}\right)^2$. By the following simple manipulations
\begin{eqnarray}
	\frac{1}{N^2} \left\llangle \left( \tr \frac{1}{p-M}\right)^2 \right\rrangle &=& \left( \frac{1}{N} \left\llangle  \tr \frac{1}{p-M}  \right\rrangle\right)^2 + \frac{1}{N^2} \left\llangle \tr \frac{1}{p-M}  \tr \frac{1}{p-M} \right\rrangle_{conn.} \nonumber\\
    &=& \W(p)^2 + \frac{1}{N^2} \W(p,p) \, ,
\end{eqnarray}
and 
\begin{equation}
	\frac{1}{N} \left\llangle \tr \left( \frac{G'(M)}{p-M}\right) \right\rrangle
    = \int_\Sigma dm\, \rho(m) \oint_C \frac{d\omega}{2\pi i} \frac{1}{\omega-m} \frac{G'(\omega)}{p-\omega}
    = \contour{\omega}{C} \W(\omega) \frac{G'(\omega)}{p-\omega} \, ,
\end{equation}
where the positively-oriented contour $C$ encloses the singularities of $\W$ but excludes the point $p$ (and possible singularities of ``$G$''), we obtain the following equation for $\W(p)$:
\begin{equation}
	\contour{\omega}{C} \frac{V'(\omega)}{p-\omega} \W(\omega) = \left(\W(p)\right)^2 + \frac{1}{N} \contour{\omega}{C} \frac{\W(\omega)}{p-\omega} \phi_z(\omega) + \frac{1}{N^2} \W(p,p).
\end{equation}
Here $\phi_z(\omega)$ is defined by
\begin{equation}
	\phi_z (\omega) \equiv \pard{}{\omega} \log \left(1+e^{\omega-z} \right) = \frac{1}{1+e^{z-\omega}} .
\end{equation}
Our problem is specialized to a Gaussian potential, $V(x)=\frac{2}{\lambda} x^2$. This is almost identical to the well-known loop equation for the Hermitian matrix model with polynomial potential $V(\omega)$, except for the $1/N$ term on the right-hand side\cite{Ambjoern1993}. Plugging in the expansion \eqref{eq:resolvent:2} we find a series of equations which can be solved iteratively in $n$. The $n=0$ equation is unaffected by the Wilson loop insertion: 
\begin{equation}
	\contour{\omega}{C} \frac{V'(\omega)}{p-\omega} \W_0(\omega) = \left(\W_0(p)\right)^2 .  \label{W0eq}
\end{equation}
For a Gaussian potential the solution is well-known (see e.g. \cite{Marino2004}): 
\begin{equation}
	\W_0(p) = \frac{2}{\lambda} \left( p-\sqrt{p^2-\lambda}\right)
\end{equation}
This has a single branch cut along $-\sqrt \lambda < p < \sqrt \lambda$. The higher order equations are 
\begin{multline} \label{eq:loopequations}
	\left\{\hat{K}-2\W_0(p)\right\} \W_n(p) = \\
\contour{\omega}{} \frac{\phi_z(\omega)}{p-\omega} \W_{n-1}(\omega)+ 
\begin{cases} 
0, & n=1\\ 
\sum_{n'=1}^{n-1} \W_{n'}(p) \; \W_{n -n'}(p) + \W_{n-2}(p,p), & n\geq 2  
\end{cases} 
\end{multline}
where we have introduced a linear operator $\hat K$, defined as in \cite{Ambjoern1993} by
\begin{equation}
	\hat{K} f(p) \equiv \contour{\omega}{C} \vprime f(\omega) .
\end{equation}
The contour $C$ is defined as before. Note that the RHS always involves correlators with smaller $n$ than the LHS. Thus one can in principle solve iteratively to obtain any $\W_n(p)$ \footnote{In \cite{Ambjoern1993}, the general iterative solution beyond leading order relies on the fact that, unlike our eq. \eqref{eq:loopequations}, the RHS there is always a rational function of $p$ (proof by induction), so by a partial fraction decomposition can be written as a sum of powers of $(p-x)^{-1}$ and $(p-y)^{-1}$, where $x$, $y$ are the endpoints of the cut. The solution $\W_n$ is thus easily expressed in terms of a set of basis functions $\chi^{(n)}(p)$, $\Psi^{(n)}(p)$, determined explicitly there, with the property that
\begin{eqnarray}
	\left\{ \hat{K} - 2\W_0(p) \right\} \chi^{(n)}(p) &=& (p-x)^{-n}\\
    \left\{ \hat{K} - 2\W_0(p) \right\} \Psi^{(n)}(p) &=& (p-y)^{-n}
\end{eqnarray}
In general the operator $\left\{\hat{K}-2\W_0(p)\right\}$ can also have zero modes; in such cases, assuming a single cut, this freedom is constrained by the large-$p$ asymptotics of $\W(p)$.
}.

In the rest of this section we shall solve \eqref{eq:loopequations} up to $n=2$.

\subsection{Solution of the loop equations}
The $n=1$ equation is 
\begin{equation} \label{W1equation}
	\left\{\hat{K}-2\W_0(p)\right\} \W_1(p) = \contour{\omega}{} \frac{\phi_z(\omega)}{p-\omega} \W_0(\omega)
\end{equation}
From now on we specialize to $V(\omega) = \frac{2}{\lambda} \omega^2$. Note that all the $p$-dependence on the RHS is in $1/(p-\omega)$. By deforming the contour $C$ to infinity (assuming $f(p)$ has no singularities outside of $C$) we find
\begin{equation} \label{eq:inf_int}
	\left\{\hat{K}-2\W_0\right\} f(p) = \mathcal{M}(p)\sqrt{p^2-\lambda}f(p) + \contour{z}{\infty} \frac{V'(z)}{z-p} f(z)
\end{equation}
where $\oint_\infty$ means we pick up the residue at infinity, and $\mathcal{M}(p)$ is given by
\begin{equation} \label{Mp}
	\mathcal{M}(p) \equiv \contour{z}{\infty} \frac{V'(z)}{(z-p)\sqrt{p^2-\lambda}} = \frac{4}{\lambda}
\end{equation}
Therefore
\begin{equation}\label{eq}
	\left\{\hat{K}-2\W_0(p)\right\} \left(\frac{1}{(p-\omega)\sqrt{p^2-\lambda }}\right) = \frac{\mathcal{M}(p)}{p-\omega}  .
\end{equation}
(The integrand in the last term of \eqref{eq:inf_int} goes like $z^{-2}$ for large $z$).
Thus \eqref{W1equation} is solved by
\begin{equation} \label{W1sol}
	\W_1(p) = \frac{\lambda/4}{\sqrt{p^2-\lambda}} \contour{\omega}{C} \frac{\phi_z(\omega) \W_0(\omega)}{(p-\omega)}
\end{equation}
Shrinking the contour to lie along the real axis, and rescaling $\omega$ so the cut extends from $-1$ to $+1$, we have
\begin{equation} \label{eq:W1sol2}
	\W_1(p) = \frac{-1}{2\sqrt{p^2-\lambda}} \int_{-1}^{1} \! d\omega \frac{\sqrt{\omega^2-\lambda}}{(p-\omega)} \frac{1}{1+e^{z-\sqrt\lambda \omega}}
\end{equation}
We now proceed to the $n=2$ equation:
\begin{equation} \label{W2}
	\left\{\hat{K}-2\W_0(p)\right\} \W_2(p) = \underbrace{\contour{\omega}{} \frac{\phi_t(\omega)}{p-\omega} \W_1(\omega)}_{a} + \underbrace{\left( \W_1(p) \right)^2\vphantom{\frac{\phi}{p}}}_{b} + \underbrace{\W_0(p,p)\vphantom{\frac{\phi}{p}}}_{c} .
\end{equation}
By linearity we can write the solution as
\begin{equation}
	\W_2(p) = \W_{2,a}(p) + \W_{2,b}(p) + \W_{2,c}(p)
\end{equation}
where $\W_{2,i}$ solves \eqref{W2} with only term $i$ on the RHS. In fact we will only need to solve for $\W_{2,a}$. We do not need $\W_{2,c}$ since the corresponding contribution to the free energy $\mathcal{F}_{2,c}$ is precisely the genus 1 ($n=2$) component of the Gaussian free energy $\mathcal{F}_{Gauss,2}$, which cancels in \eqref{eq:Fz}. Nor is $\W_{2,b}$ relevant, because the asymptotics $\W_1(p\rightarrow\infty) \sim p^{-2}$ imply that it does not contribute to $\partial_\lambda \mathcal{F} \propto \contour{p}{\infty} p^2 \, \W(p)$ \footnote{
It is however easy to show using \eqref{eq:inf_int} that $\W_{2,a}$ is just
\begin{equation}
	\W_{2,b}(p) =\frac{\lambda/4}{\sqrt{p^2-\lambda}} \left(\W_1(p)\right)^2
\end{equation}
}.

Here the $p$-dependence of the RHS is the same as for the $n=1$ equation and the solution is therefore analogous:
\begin{eqnarray}
	\W_{2,a}(p) &=& \frac{\lambda/4}{\sqrt{p^2-\lambda}}\contour{\omega}{C} \frac{\phi(\omega) \W_1(\omega)}{p-\omega} \\
    &=& \frac{(\lambda/4)^2}{\sqrt{p^2-\lambda}} \contour{u}{C'} \frac{\phi(u)}{p-u}\frac{1}{\sqrt{u^2-\lambda}} \contour{v}{C} \frac{\phi(v)}{u-v} \W_0(v)\\
\end{eqnarray}
The contour $C'$ encloses $C$ but not $p$. Only the singular (square root) part of $\W_0(p)$ contributes. Again shrinking $C$, $C'$ around the cut, we can write
\begin{equation} \label{eq:W2a}
	\W_{2,a}(p) = \frac{-\lambda}{8\pi^2\sqrt{p^2-\lambda}} \int_{-\sqrt{\lambda}}^{\sqrt{\lambda}}  du \, \frac{\phi(u)}{(p-u)\sqrt{\lambda -u^2}} \strokedint_{-\sqrt{\lambda}}^{\sqrt{\lambda}} dv\, \frac{\phi(v) \sqrt{\lambda-v^2}}{u-v}
\end{equation}
where $\strokedint$ denotes the Cauchy principal value.

\subsection{Free energy I: \texorpdfstring{$\mathcal F$}{F} as generator of \texorpdfstring{$\langle (\tr M^2)^l \rangle $}{<trM2>}}
The resolvent \eqref{eq:resolvent:1} is the generator of  expectation values of monomials
\begin{equation}
	\W(p)=\frac{1}{N}\sum_{k=0}^{\infty} \frac{\llangle \tr M^k  \rrangle }{p^{k+1}} .
\end{equation}
This allows us to obtain the free energy as an integral over $\lambda$ and $p$, since
\begin{eqnarray}
	\frac{\lambda^2}{2N^2} \partial_\lambda \log \mathcal{Z}(z) &=& \frac{1}{N} \left\llangle \tr M^2 \right\rrangle \\
    &=& \contour{p}{\infty} p^2 \W(p)
\end{eqnarray}
The potential $V(M) = \frac{2}{\lambda} M^2$ acts as a source for insertions of $\tr M^2$. Thus we have
\begin{equation}
	\mathcal{F}(z) = -2 N \int  \frac{d\lambda}{\lambda^2} \contour{p}{\infty} p^2\W(p) +C_0(z) -\mathcal{F}_{Gauss}
\end{equation}
with some integration constant $C_0(z)$. Due to the subtraction of $\mathcal F_{Gauss}$, the leading order is determined by $\W_1$. The $p$-integral is easily done
\begin{equation} \label{ResAtInfinity}
	\underset{p=\infty}{\text{Res}} \; \frac{p^2}{(p-\omega)\sqrt{p^2-\lambda}} = -\omega,
\end{equation}
as is the $\lambda$ integral, and we find
\begin{equation}
	\mathcal{F}_0(z) = -\frac{2N}{\pi}\int_{-1}^{+1} \!\! d\omega \,\sqrt{1-\omega^2}\; \log \left( 1+ e^{\sqrt{\lambda}\,\omega-z}\right) + C_0(z)
\end{equation}
in agreement with \eqref{eq:planarresult}. Proceeding in the same way with $\W_{2,a}$, we get
\begin{equation}
	\mathcal{F}_{1}(z) = \frac{-1}{4\pi^2}\int\!\!d\lambda \int_{-1}^1 \!\!du \strokedint_{-1}^1  \!\!dv\, \frac{u}{\sqrt{1-u^2}} \, \frac{\sqrt{1-v^2}}{u-v}\, \frac{1}{1+e^{z-\sqrt{\lambda}u}}\, \frac{1}{1+e^{z-\sqrt{\lambda}v}}
\end{equation}
Recall that $z$ here is to be substituted with $z_* = \sqrt \lambda \cos \theta_k $. At strong coupling the Fermi-Dirac function can be approximated by a step function - this is the first term in the ``low-temperature'' expansion. Thus
\begin{equation}
	\partial_\lambda \mathcal{F}_1(z) \simeq \frac{-1}{4\pi^2} \int_\zeta^1 du \frac{u}{\sqrt{1-u^2}} \strokedint_\zeta^1 dv\, \frac{\sqrt{1-v^2}}{u-v}
\end{equation}
where $\zeta \equiv z_*/\sqrt \lambda$. We find for the $v$-integral, taking care with the principal value, that
\begin{equation} \label{eq:principalvalueint}
	\frac{1}{\sqrt{1-u^2}} \strokedint_\z^1 \! dv \, \frac{\sqrt{1-v^2}}{u-v} = \frac{u \arccos(\z) +\sqrt{1-\z^2}}{\sqrt{1-u^2}} - \log \frac{1-u\z + \sqrt{1-u^2}\sqrt{1-\z^2}}{|u-\z|}.
\end{equation}
For the $u$ integral we get
\begin{equation}
	\int_\z^1 du\, \frac{a u^2 + b u}{\sqrt{1-u^2}} = \frac{1}{2} \arccos^2(\z) + (1-\z^2) + \frac{1}{2} \z\sqrt{1-\z^2} \arccos(\z)
\end{equation}
and
\begin{equation}
	\int_\z^1 \!du\, u\, \log \frac{1-u\z + \sqrt{(1-\z^2)(1-u^2)}}{u-\z} 
    = \frac{1}{2}\left[  1-\z^2+\z\sqrt{1-\z^2} \arccos(\z)
    \right]
\end{equation}
resulting in
\begin{equation}
	\partial_\lambda \mathcal{F}_1(z) =  \frac{-1}{8\pi^2} \left[ \arccos^2(z_*/\lambda) + (1-z_*^2/\lambda)\right]  .
\end{equation}
Finally this can be integrated with the help of Mathematica to give
\begin{equation}
	\int^\lambda \! d\lambda \, \left[ \partial_\lambda \mathcal{F}_1(z) \right] = \frac{-1}{8\pi^2} \left[ \lambda -2 z \sqrt{\lambda-z^2} \arccos \left(\frac{z}{\sqrt{\lambda }}\right)+\lambda  \arccos^2\left(\frac{z}{\sqrt{\lambda }}\right)\right] + C_1(z).
\end{equation}
In terms of the $\sqrt \lambda$-scaled parameter we then have
\begin{subequations} \label{eq:Fint}
\begin{equation} \label{eq:Fint:1}
	\mathcal F _1(z) = -\frac{\lambda}{8\pi^2} \left[ 1-2z\sqrt{1-z^2} \arccos(z) + \arccos^2(z) \right] + C_1(\sqrt \lambda z)
\end{equation}
What about the integration constant $C_1(z)$? The leading $\lambda$-dependence obtained here suggests it should be of the form $C_1(z) = a z^2$. Then the requirement that $\mathcal{F}_1 = 0$ at $z=1$ (corresponding to $k=\theta=0$) fixes $a$ to be $1/8\pi^2$:
\begin{equation} \label{eq:Fint:2}
	C_1(z) = \frac{z^2}{8\pi^2}
\end{equation}
In terms of $\theta_k$ the result is
\begin{equation} \label{eq:Fint:3}
	\mathcal F_1(\theta_k) =  -\frac{\lambda}{8\pi^2} \left[ \sin^2\theta_k-\theta_k \sin 2\theta_k + \theta_k^2 \right] 
\end{equation}
\end{subequations}
In the next section we will show that with this choice of $C(z)$, $\mathcal F_1$ agrees with the direct evaluation of the matrix integral using the eigenvalue density.

\subsection{Free energy II: \texorpdfstring{$\rho_n$}{rho} from \texorpdfstring{$\W_n(p)$}{Wn}}
An alternative route to the free energy is via the eigenvalue density. The first $1/N$ correction, $\mathcal F_1$, will require knowledge of the backreacted eigenvalue density, which is encoded in the resolvent\footnote{Fluctuations 
around the large-$N$ eigenvalue saddlepoint, which would contribute a factor of $\frac{1}{2}\log \det [ \partial^2 S/\partial m_i \partial m_j]$ to the free energy, do not contribute at this order as they cancel against the equivalent contribution to $\mathcal{F}_{Gauss}$.}.

\begin{figure}[tbp]
\centering 
\includegraphics[width=.75\textwidth]{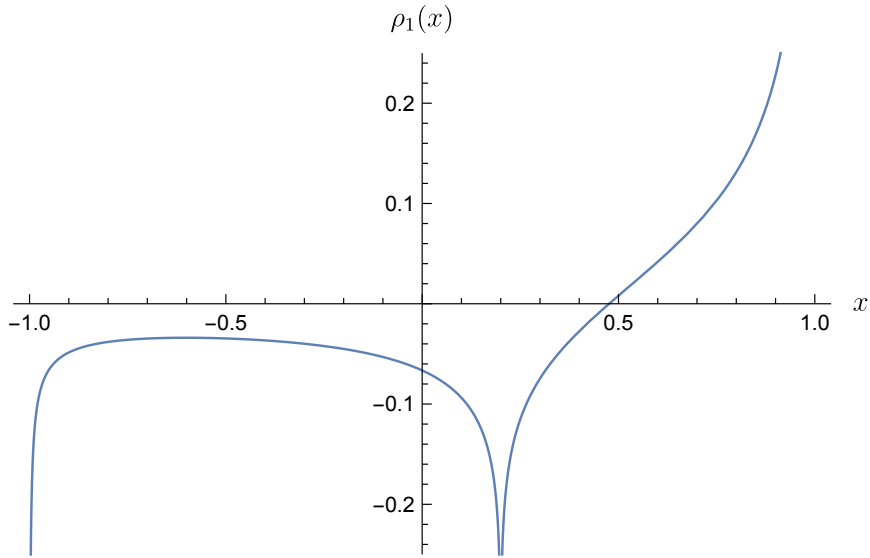}
\caption{\label{fig:density} $\rho_1(\sqrt \lambda x)$, the $1/N$ correction to the density, at large $\lambda$ and with $z=0.2$. The analytic expression is given by \eqref{eq:density}.}
\end{figure}

The action \eqref{eq:action} in terms of $\rho$ is
\begin{eqnarray}
	S_0 &=& \frac{2}{\lambda} \int \! dx \, \rho (x) x^2 - 2 \iint  \! dx dy\, \rho(x) \rho (y) \log |x-y| \\
    S_1 &=& -\int\!dx\, \rho(x) \log(1+ e^{x-z})
\end{eqnarray}
Expanding around $\rho_0$ gives
\begin{multline}\label{expansionofS}
	S\left[\rho_0 + \rho_1/N + \mathcal O(1/N^2)\right] = S_0[\rho_0] + \frac{1}{N} \left\{ \int \rho_1 \left.\frac{\d S_0}{\d \rho}\right|_{\rho_0}  + S_1[\rho_0] \right\} \\
    + \frac{1}{N^2} \left\{ \frac{1}{2} \iint \rho_1 \frac{\d^2 S_0}{\d \rho \d \rho } \rho_1 + \int \rho_1 \left.\frac{\d S_1}{\d \rho}\right|_{\rho_0} + \int \rho_2 \left.\frac{\d S_0}{\d \rho}\right|_{\rho_0} \right\} + \mathcal O(\frac{1}{N^3})
\end{multline}
The terms involving first derivatives of $S_0$ are identically zero, by the equation of motion. We can also eliminate the awkward double integral by means of the $\mathcal O(1/N)$ equation of motion:
\begin{equation}
	0 = \deriv{}{x} \left[ \int \frac{\d^2 S_0}{\d \rho(x) \d \rho(y) } \rho_1(y) dy  +  \left.\frac{\d S_1}{\d \rho}\right|_{\rho_0} \right]
\end{equation}
Thus integrating the quantity in square brackets against $\rho_1(x)$ (a trick used in \cite{Chen-Lin2016nonplanarSymmWL}) gives
\begin{equation}
	\iint \rho_1 \frac{\d^2 S_0}{\d \rho \d \rho } \rho_1 = -\int \rho_1 \left. \frac{\d S_1}{\d \rho}\right|_{\rho_0}
\end{equation}
Therefore the action finally reduces to 
\begin{equation}
	N^2 \,S[\rho] = N^2 S_0[\rho_0] - N \int\! dx\, \rho_0(x) \log(1+e^{x-z}) - \frac{1}{2} \int \!dx\, \rho_1(x) \log(1+e^{x-z})
\end{equation}
The first term is canceled by $\mathcal F_{Gauss}$, the second is the planar result in \eqref{eq:planarresult}, and the third is the one we are after. After scaling of $x$, $z$ by $\sqrt \lambda$ as before, we have
\begin{subequations}
\begin{eqnarray} \label{eq:f1rho}
	\mathcal F_1(z) &=& -\frac{\sqrt{\lambda}}{2} \int_{-1}^{1} \! dx \, \rho_1(\sqrt\lambda x) \log (1+ e^{\sqrt\lambda (x-z)})  \\  \label{eq:f1rho:1}
    &\simeq &  - \frac{\lambda}{2} \int_{z}^{1} \! dx \, \rho_1(\sqrt\lambda x) (x-z), \qquad (\lambda\rightarrow \infty)  \label{eq:f1rho:2}
\end{eqnarray}
\end{subequations}
We now determine $\rho_1(\sqrt\lambda x)$ from $\W_1(p)$. Recall that the continuum form of \eqref{eq:resolvent:1}, namely $\W(p) = \int \frac{\rho(m)}{p-m} \, dm$, implies that the eigenvalue density is given as the discontinuity across the cut:
\begin{equation} \label{densityfromW}
	\rho(x) = \frac{1}{2\pi i} \left( \W(x-i\epsilon) -\W(x+i\epsilon) \right)
\end{equation}
Using the relation
\begin{equation}
	\frac{1}{x\pm i\epsilon} = \mathcal{P} \left( \frac{1}{x} \right) \mp i \pi \delta (x) ,
\end{equation}
where $\mathcal{P}$ denotes the Cauchy principal value,
and recalling \eqref{eq:W1sol2}, which we repeat here,
\begin{equation*}
	\W_1(p) = \frac{-1}{2\sqrt{p^2-\lambda}} \int_{-1}^{1} \! d\omega \frac{\sqrt{\omega^2-\lambda}}{(p-\omega)} \frac{1}{1+e^{z-\sqrt\lambda \omega}} ,
\end{equation*}
we find
\begin{equation} \label{eq:rho1}
	\rho_1(x) = \frac{1}{2\pi^2 \sqrt{\lambda-x^2}} \strokedint_{-\sqrt{\lambda}}^{\sqrt{\lambda}} d\omega\, \phi(\omega) \frac{\sqrt{\lambda-\omega^2}}{(x-\omega) }
\end{equation}
As usual we re-scale $\omega$ and $z$ by $\sqrt{\lambda}$. Then at strong coupling we can approximate $\phi_z(\sqrt{\lambda}\omega) \simeq \theta (\omega-z)$, i.e.
\begin{equation}
	\rho_1(\sqrt{\lambda} x) \approx \frac{1}{2\pi^2 \sqrt{1-x^2}} \strokedint_{z}^{1} \!d\omega\, \frac{\sqrt{1-\omega^2}}{(x-\omega) }
\end{equation}
Using \eqref{eq:principalvalueint} we thus find
\begin{equation} \label{eq:density}
	\rho_1(\sqrt{\lambda} x) = \frac{1}{2\pi^2} \left\{ \frac{x \arccos(z) +\sqrt{1-z^2}}{\sqrt{1-x^2}} - \log \frac{1-x z + \sqrt{1-x^2}\sqrt{1-z^2}}{|x-z|}  \right\}
\end{equation}
This function is plotted in fig. \ref{fig:density}. Note the logarithmic singularity that arises at infinite coupling, located on the cut at $x = z$. (At finite $\lambda$, \eqref{eq:rho1} is a smooth function of $x$). $\rho_1(x)$ is correctly normalized to zero: $\int_{-1}^1 dx \rho_1(\sqrt{\lambda} x) = 0$.

The free energy (at strong coupling) now follows from \eqref{eq:f1rho} and \eqref{eq:density}. The result is
\begin{subequations} \label{eq:F1un}
\begin{equation} \label{eq:F1un:z}
	\mathcal{F}_1(z) = \frac{-\lambda}{8\pi^2} \left(1-z^2 - 2z \sqrt{1-z^2} \; \arccos(z)+\arccos(z)^2\right) ,
\end{equation}
or in terms of $\theta_k$:
\begin{equation} \label{eq:F1un:theta}
	\mathcal{F}_1(\theta) = \frac{-\lambda}{8\pi^2} \left( \sin^2\theta - \theta \sin 2\theta + \theta^2 \right),
\end{equation}
\end{subequations}
in precise agreement with \eqref{eq:Fint}.

\subsection{Modification for \texorpdfstring{$SU(N)$}{SU(N)}}
AdS/CFT is generally held to describe $\mathcal N = 4$ SYM with gauge group $SU(N)$. This is motivated by considering the Kaluza-Klein spectrum of IIB supergravity. On the CFT side, for a $U(N)$ gauge theory, the $U(1)$ and $SU(N)$ components decouple, up to global identifications. On the AdS side, dimensional reduction of SUGRA on the internal $S^5$ does indeed give rise to a free $U(1)$ multiplet, but this comprises pure gauge modes which can be set to zero in the bulk (see e.g. \cite{Aharony2000,Zaffaroni2000}). 

Since we are studying $1/N$ effects here, the difference between the $U(N)$ and $SU(N)$ theories is potentially important; so far we have only considered the former. For $SU(N)$ the integral \eqref{eq:HermitianMatrixModel}  is over traceless Hermitian matrices. It is not hard to integrate out the trace degree of freedom explicitly\footnote{
Alternatively we can keep the integral over all of $\mathfrak u(N)$ and impose the tracelessness constraint with a Lagrange multiplier $\Lambda$. This adds $N^2 S_\Lambda = \Lambda \, \sum_{i=1}^N m_i $ to the action, resulting in a perturbation $\delta \rho_\Lambda (\sqrt \lambda x) = \frac{\Lambda}{2\pi} \frac{x}{\sqrt{1-x^2}}$ to the density. The tracelessness condition $\int_{-1}^1\! dx \, x\, \left[\rho_1(x) + \delta\rho_\Lambda (x)\right] = 0$ then fixes the multiplier to $\pi \Lambda = z \sqrt{1-z^2}-\arccos(z) $. Using the saddlepoint equation \eqref{eq:saddlepointeqn} for $z$ (or $\theta_k$), we then find precisely the result \eqref{eq:UN_SUN} above.
}
, as described for the fundamental case in \cite{Drukker2001}. Write $M = M' + mI$, where $M'$ is traceless. The measure is just $[dM]=[dM']dm$. From the definition of $\WA$ in terms of the generating function \eqref{eq:generatingfn} we have
\begin{equation}
	\ev{\WA}_{U(N)} = \left\langle e^{k m} \partial_{\tilde t}^{N-k} \det(\tilde t + e^{M'})\right\rangle_0 \,\bigg|_{\tilde t=0}
\end{equation}
where we made the replacement $\tilde t = t e^{-m}$. Integrating out $m$ we obtain the following exact relation between the Wilson loops of the two theories:
\begin{equation} \label{eq:UN_SUN}
	\ev{\WA}_{U(N)} = e^{k^2 \lambda/8N^2} \left\langle \WA \right\rangle_{SU(N)}
\end{equation}
In terms of the free energies we have $\mathcal{F}_{SU(N)}(z) =\mathcal{F}_{U(N)}(z) + \frac{1}{8} \lambda f^2$. Our final result for $\mathcal F_1(\theta_k)$ is remarkably simple:
\begin{equation} \label{eq:F1sun}
	\mathcal{F}_1^{SU(N)} = -\frac{\lambda}{8\pi^2} \, \sin^4 \theta .
\end{equation}



\section{Numerical checks} \label{sec:numerics}


The antisymmetric Wilson loop was evaluated in \cite{Fiol2014a} using orthogonal polynomials, yielding the following \emph{exact} result for the generating function \eqref{eq:generatingfn} of the Wilson loop
\begin{equation} \label{eq:exactresult}
	F_\mathcal{A}(t) = \det \left[ t + A\, e^{\lambda/8N}  \right], \qquad A_{ij} \equiv L_{j-1}^{i-j} (-\lambda/4N) ,
\end{equation}
which we evaluate numerically for large values of $N$ and $\lambda$. In order to compare this with our $\mathcal{F}_1$ we must subtract off the planar result at strong coupling. As detailed in \cite{Horikoshi2016}, the latter is obtained as an expansion in large-$\lambda$ using the ``low-temperature'' expansion of the Fermi-Dirac function which appears in the planar saddle-point equation. We find
\begin{multline} \label{eq:planarexpansion}
	\left( \log \WA \right)_{\textrm{planar}} = \frac{4\pi N}{\lambda}
 \left[\lambda^{\frac{3}{2}} \frac{\sin^3\theta_k}{6\pi^2}
 +\sqrt{\lambda}\,\frac{\sin\theta_k}{12}
-\frac{1}{\sqrt \lambda} \frac{\pi^2}{1440} \frac{(19+5\cos 2\theta_k)}{ \sin^3\theta_k} \right.
\\
\left. - \frac{1}{\lambda^{\frac{3}{2}}} \frac{\pi^4}{725760} \frac{ (6788 \cos 2\theta_k + 35 \cos 4 \theta_k + 8985)}{\sin^7\theta_k}
+\cdots\right] ,
\end{multline}
where we have corrected a numerical error in the last two terms of eq. 2.10 of \cite{Horikoshi2016}\footnote{I thank Kazumi Okuyama for correspondence on this point.}.

\begin{figure}[tbp]
\centering 
\includegraphics[width=.49\textwidth]{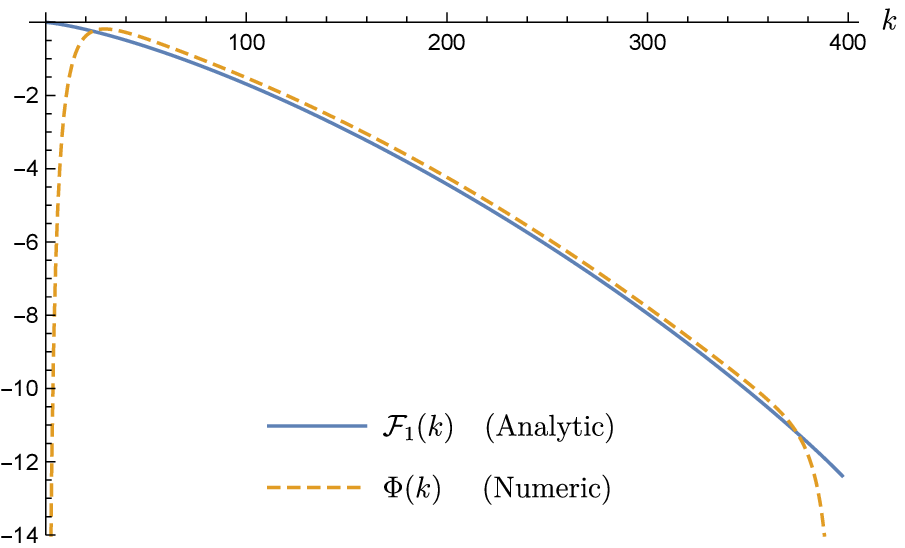}
\hfill
\includegraphics[width=.49\textwidth,origin=c]{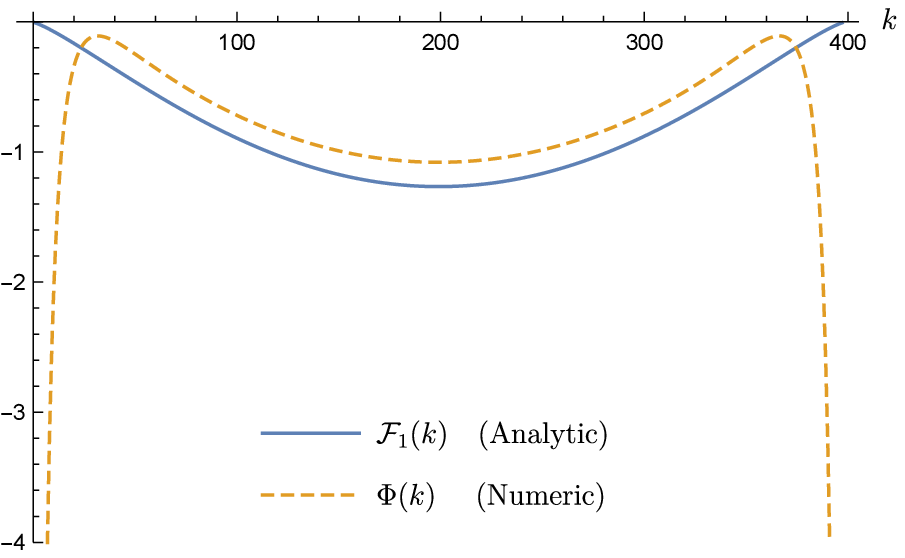}
\caption{\label{fig:NumericalComparison} Numerical versus analytic results for $\mathcal F_1$ as a function of $k$. This was defined via equations (\ref{eq:Fz},\ref{eq:Fexpansion},\ref{eq:saddlepointeqn}). The solid blue line is our analytic result for $\mathcal F_1$, and its numerical approximation $\Phi(k)$ (defined in \eqref{eq:Phi}) is given by the orange dashed line. The plot on the left is for gauge group $U(N)$ (eq. \eqref{eq:F1un}) while that on the right is for $SU(N)$ (eq. \eqref{eq:F1sun}). (As in figure \ref{fig:strongcoupling}, we have replaced $d_A =\binom{N}{k} \rightarrow 1$ here).}
\end{figure}

Finally, to compare precisely with the numerics, we need all other contributions up to $\mathcal O (N^0)$. This includes the prefactor $\binom{N}{k}^{-1} \frac{\sqrt{\lambda }}{2 \pi }$ from \eqref{eq:WLcontour}, as well as the 1-loop contribution from the $z$-integral:
\begin{equation}
	\sqrt{\frac{2\pi}{N \mathcal{F}_0''(z_*)}} = \sqrt{\frac{\pi ^2}{N \sqrt{\lambda } \sin \theta_k} }.
\end{equation}
With these factors included, we find good numerical agreement with the exact result \eqref{eq:exactresult}. In figure \ref{fig:NumericalComparison} we plot $\mathcal{F}_1(\theta_k)$ versus $\Phi(k)$, defined as
\begin{equation} \label{eq:Phi}
	\Phi(k) \equiv -\log (\WA^{\textrm{exact}}) + \left( \log \WA \right)_{\textrm{planar},3} + \frac{1}{2} \log \frac{\sqrt\lambda}{4N d_A^2}
    + \begin{cases}
    \phantom{\frac{\lambda}{8}}0 \hspace{\stretch{1}}, & U(N) \\
    \frac{\lambda}{8} \left(\frac{k}{N}\right)^2 ,  & SU(N)
    \end{cases}
\end{equation}
where $\left( \log \WA \right)_{\textrm{planar},3}$ contains the first three terms of the planar strong coupling expansion \eqref{eq:planarexpansion}. 
The parameter values used are $N=400$, $\lambda=100$. We then expect the next correction to to be $\mathcal O(10^{-1})$, since for the ``higher-genus'' and strong-coupling corrections we have respectively $\left(\frac{\sqrt \lambda}{N}\right)^2 \log \WA \approx 0.5$ and $\left(\frac{\sqrt \lambda}{N}\right)\frac{1}{\lambda} \log \WA \approx 0.2$.
This is indeed borne out by the numerics: from the plot we see that the residual is approximately 
$
\left|  \mathcal F_1^{numeric}-\mathcal{F}_1^{analytic}  \right|  \approx 0.2 .
$
If instead we take $N=700$, $\lambda = 30$, so that $\frac{1}{\lambda} \gg \frac{\sqrt \lambda}{N}$, we get 
$$\left|  \mathcal F_1^{numeric}-\mathcal{F}_1^{analytic}  \right| \approx\left(\frac{\sqrt \lambda}{N}\right)\frac{1}{\lambda} \log \WA \approx 0.25,
$$
whereas $\left(\frac{\sqrt \lambda}{N}\right)^2 \log \WA \approx 0.05$. 

\begin{figure}
\centering
\begin{tikzpicture}
\node (img1)  {\includegraphics[width=.5\textwidth]{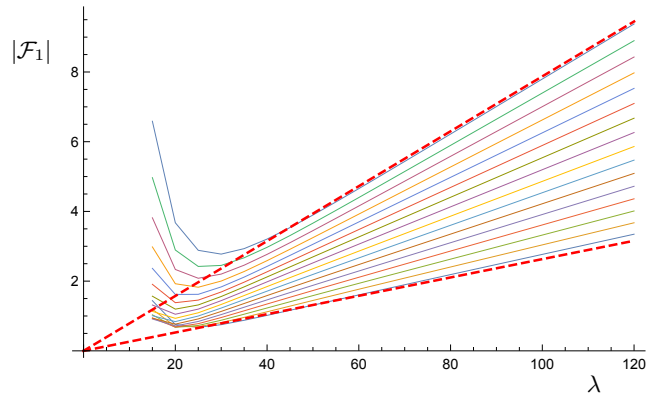}};
  \node[below of=img1, node distance=0cm, yshift=-2.6cm,xshift=3.1cm] {\footnotesize$\lambda$};
  \node[left of=img1, node distance=0cm, 
  anchor=center,yshift=1.8cm,xshift=-4.3cm] {\footnotesize$\left| \mathcal F_1 \right|$};
\end{tikzpicture}
\caption{\label{fig:lamdep} Linear dependence of $\mathcal F_1(k)$ on $\lambda$: Each line corresponds to a particular value of $k$ in the range $100<k<300$, with $N=400$. The dashed rays are included simply as visual aids. For large enough $\lambda$ we see precisely the linear behavior obtained in \eqref{eq:F1un}.}
\end{figure}

Finally, as a check of the $\lambda$ dependence, in figure \ref{fig:lamdep} we plot $\mathcal F_1^{numeric}(k)$ versus $\lambda$, for fixed $N$ and several different values of $k$. This plot clearly illustrates the linear behavior at large $\lambda$.


\section{Conclusions} \label{sec:Conclusions}
We computed the first $1/N$ correction to the $\frac{1}{2}$-BPS circular antisymmetric Wilson loop of rank $k$, with $k$ of order $\mathcal{O}(N)$, in $\mathcal N=4$ SYM, at leading order in 't Hooft coupling $\lambda$. The result is given in equations \eqref{eq:F1un} and \eqref{eq:F1sun}, for gauge group $U(N)$ and $SU(N)$ respectively.


The holographic dual of this object is a known probe $D5$-brane configuration with $k$ units of electric flux on its worldvolume. Interestingly, the results obtained here and the $D$-brane $1$-loop effective action computed in \cite{Faraggi2012OneLoopEffectiveActionAntisymmetricWL} do not match. There they found $\mathcal F_1 \sim \mathcal{O}(N^0 \lambda^0)$. In contrast, our calculation yielded $\mathcal{F}_1 \sim \mathcal{O}(\lambda)$, implying an expansion in $\sqrt{\lambda}/N$. The obvious explanation for this discrepancy is the gravitational backreaction of the brane, which so far has not been accounted for. Integrating out the bulk action in the Gaussian approximation would indeed give a result $\mathcal O(\lambda)$, although the problem is not so simple as one needs to account for the  infinite tower of Kaluza-Klein modes and their couplings to the brane\footnote{I thank Kostya Zarembo for comments on this point.}. It also remains desirable to resolve the numerical mismatch at $\mathcal{O} (N^0,\lambda^0)$, by studying $1/\lambda$ corrections.

It is worth mentioning that several important properties of the heavy probes studied here follow directly from the Wilson loop expectation value, including the so-called Bremsstrahlung function\footnote{see also \cite{Bianchi2014,Bianchi2017,Bianchi2017a} for a similar formula in the context of ABJM theory} \cite{Correa2012a}
\begin{equation}
	B_{\mathcal A_k}(\lambda, N) = \frac{1}{2\pi^2} \lambda \partial_\lambda \ev{W_{\mathcal{A}_k}} ,
\end{equation}
and the additional entanglement entropy $\Delta \mathcal S$, relative to the vacuum, of a spherical region threaded by the probe\footnote{
Incidentally, it should be possible to calculate $\Delta \mathcal S $ holographically using the approach of  \cite{Chang2014b}, whose authors studied the additional holographic entanglement entropy due to the presence of probe branes. The leading order effect arises from the backreaction of the probes on the geometry, and the concomitant distortion of the Ryu-Takayanagi minimal surface. This was shown to be captured by a compact ``double-integral'' formula, where the integrations are taken over the brane worldvolume and unperturbed minimal surface respectively, obviating the need for a full, backreacted solution. As argued in some detail in \cite{Chang2014b}, complications due to fields other than the metric being sourced by the brane may be avoided, thanks to the particular worldvolume gauge field configuration relevant to this problem.
} 
\cite{Lewkowycz2014,Gentle2014},
\begin{equation}
	\Delta \mathcal S = \left( 1- \frac{4}{3} \lambda \partial_\lambda \right) \log \ev{W_{\mathcal{A}_k}}. 
\end{equation}

Also intriguing is the relation of the antisymmetric Wilson loop to a supersymmetric Kondo model \cite{Harrison2012}.

The plethora of gauge theory localization results  in the literature opens the door to a number of natural extensions of the present work. Firstly, there exist exact results for various gauge theories with generally richer structure than the highly symmetric $\mathcal N=4$ SYM. Expectation values of higher rank SUSY Wilson loops have  been studied in the planar limit in $\mathcal N=2^* $ SYM \cite{Chen-Lin2015HigherRankWLinNeq2star}, $\mathcal N=2$ SQCD \cite{Fraser2012} and also ABJM theory \cite{Cookmeyer2016HigherRankABJM_WLs}\footnote{In the latter work, a partial $1/N$ contribution, analogous to the logarithmic term in \eqref{eq:logWexpansion}, was also calculated.}.  On the gravity side some of the corresponding probes have been studied in eg. \cite{Chen-Lin2016HologSymmWL,Faraggi2012OneLoopEffectiveActionAntisymmetricWL,Faraggi2011,Mueck2016}. The problem of 1-loop matching remains open in all these cases.

Continuing in this vein, we could also consider more general correlators, again beyond the planar limit. The resolvent \eqref{eq:resolvent}, which we have obtained from the loop equation, encodes expectation values of monomials in the presence of the Wilson loop, $\langle \tr M^j \rangle_{\WA}$. For the Hermitian matrix model these have no direct physical interpretation. However, the analogous quantities in the normal matrix model describe correlators of the Wilson loop with chiral primary operators in $\mathcal N=4$ SYM \cite{Okuyama2006,Giombi2006}, and it would be interesting to extend our analysis to this case. On the gravity side, the corresponding ``backreaction'' calculation may prove more tractable than that of the Wilson loop expectation value itself. 

This story generalizes still further to a larger subsector of Wilson loops and chiral primary operators in $\mathcal N = 4$ SYM. For example, one can consider the generically \mbox{$\frac{1}{8}$-BPS} configurations of multiple loops and chiral primaries supported on an $S^2$ submanifold of $\mathbb R^4$. It is believed that correlators of such observables reduce to bosonic 2d Yang-Mills theory \cite{Drukker2007,Drukker2008,Drukker2008a,Pestun2012,Giombi2013}, which in turn can be mapped to certain multi-matrix models. (This is still at the level of conjecture, as the 1-loop fluctuations around the localization locus have not been explicitly evaluated. See \cite{Bassetto2008,Bassetto2009,Bonini2014} however for several non-trivial checks of the conjecture). Aspects of the matrix model machinery we have employed can be generalized to the study of multi-matrix models.






\acknowledgments

The author would like to thank Gordon Semenoff and Konstantin Zarembo for useful discussions, as well as Xinyi Chen-Lin for early conversations on this topic. This work was supported by the Marie Curie network GATIS of the European Union's FP7 Programme under REA Grant Agreement No 317089.

\bibliographystyle{utphys}
\bibliography{wilsonlooprefs} 









\end{document}